\documentstyle[aps,twocolumn,psfig]{revtex}
\newcommand{\el}{{\it el al.},}
\begin{document}
\draft
%\twocolumn
\twocolumn[\hsize\textwidth\columnwidth\hsize\csname @twocolumnfalse\endcsname
\title{Surface states and their possible role in the superconductivity of MgB$_2$}
\author{V.D.P.\ Servedio$^1$, S.-L.\ Drechsler$^2$, T.\ Mishonov$^3$ }
\address{$^1$Institut f\"ur Oberfl\"achenphysik und Mikrostrukturphysik,
TU Dresden, D-01062 Dresden, Germany}
\address{$^2$Institut f\"ur Festk\"orper- und Werkstofforschung Dresden, D-01171 Dresden,
Germany}
\address{$^3$LVSM, Katholieke Universiteit, Leuven, Belgium and
University of Sofia, Bulgaria}
%\date{\today}
\maketitle
%\begin{center}For {\em Phys. Rev. X} \end{center}
%\widetext
 
%***************** VM screen width ************************************%
\begin{abstract}

We report layer-Korringa-Kohn-Rostocker 
(LKKR) calculations for bulk and surface states (SFS) 
as well as the corresponding 
photoemission intensities of MgB$_2$. Our theoretical results 
reproduce very well the recent angle resolved photoemission data by 
Uchiyama {\it et al}, cond-mat/0111152. At least two SFS 
are assigned.
Consequences of SFS on the anisotropy of the 
upper critical fields and other properties in the superconducting state 
of grains in micro-powder samples are discussed. 
\end{abstract}

] % ends twocolumn:FALSE
 
%\vfill\eject
% body of paper here
%\twocolumn
%\section{Introduction}
Recently, the first angle-resolved photoemission spectroscopy (ARPES)
measurements on a MgB$_2$ single crystal were reported 
by Uchiyama {\it et al.} \cite{tajima}.
They observed several dispersive states. Among them one  has been
ascribed to a surface state (SFS). The very existence 
of SFS and subsurface states for MgB$_2$ as well as  
surface superconductivity 
have been predicted first in Refs.~2 and~3.
However, there are significant quantitative differences between Ref.\ 2 
and the ARPES data. 
The effects of SFS for the superconductivity in this challenging
compound have not yet been discussed in detail.
In particular, there is an unresolved puzzle with respect to the
out-of plane anisotropy for the upper critical fields 
$\gamma_H=H^{ab}_{c2}/H^c_{c2}$. 
It is moderate for  single crystals $1.6 \leq\! \gamma_H\! \leq 3.5$. 
\cite{lima} 
But large values $6 \leq\! \gamma_H\! \leq 9$
have been deduced indirectly from  measurements 
on pure powder samples with small grains.   
\cite{budkoaniso,simon,papavassiliou}
Enhanced superconductivity due to locally enhanced DOS (density of states)
has been predicted \cite{kim} 
for idealized B- as well as Mg-terminated surfaces (BTS, MgTS) although in
 conflict 
with experimental data. 
We report on LKKR calculations for MgTS and BTS and compare them with 
the ARPES data.\cite{tajima} 
Furthermore we propose a scenario which involves essentially SFS to resolve 
some puzzles of general interest mentioned above.

%%%%\section{Electronic structure}
The electronic DOS for the semi-infinite system
MgB$_2$ as well as the photoemission spectra, were calculated using a
LKKR code \cite{samed}.
This code treats the photoemission process within the ``one-step''
approach.
Dealing with the Green function formalism, it is capable to
work with a complex potential, allowing the description of
broadening effects due to the finite lifetime of quasi-particles.
However, a set of input parameters must be provided phenomenologically
in order to match experimental data as close as possible. 
%The experimental ARPES data available so far which 
%we referred to, are those in Ref.~\onlinecite{tajima}.
As an input to the LKKR code, the real part of the one-electron
effective complex 
potential was calculated self-consistently with the help of a
LMTO (linear muffin-tin orbital) code. 
This preliminary calculation
was done for the infinite MgB$_2$ crystal with a basal plane 
lattice constant of 
$a$=3.085~\AA\ and a ratio $c/a$=1.142. The von~Barth-Hedin \cite{BH}
correlation energy functional was used in connection with the local
density approximation (LDA). 
The resulting Fermi energy $E_F\approx$~13.1~eV measured from
 the muffin-tin zero,  while the muffin-tin radii were 1.711~\AA\ and
0.988~\AA\  for the Mg and B, respectively.
A direct evaluation of the self-energy correction to be added to the
LMTO potential is notoriously very difficult. Hence, we 
fixed it empirically.
As for its imaginary part we chose  different values for states below
and above the Fermi level, -0.05~eV and -1~eV respectively. Its
real part was chosen to be -1~eV and applied to the photo-electron
final states only.
Rigorously,  the full complex self-energy
correction is not constant in the energy intervals under consideration, 
but here such details can be omitted without losing much accuracy.
 
The geometry of the semi-infinite MgB$_2$ system was simply modeled 
cutting the ideal crystal perpendicularly to the $\hat{c}$ axis. Hence, no
possible relaxation or reconstruction effects were taken into account.
The topmost surface layer consists either of Mg atoms or B
atoms. In the MgTS-case a work function $\phi=$4.2~eV was used,
in the BTS $\phi=$6.1~eV, as suggested in Ref.~\onlinecite{silkin}.
In both cases the surface potential barrier was modeled as a 
step potential, reflecting for occupied states and non reflecting, but
refracting, for unoccupied ones. Its position $d$ was fixed to
0.987~\AA\ with respect to the topmost layer, while its height is
given by the sum $E_F+\phi$. 
The energy of impinging photons  at 28~eV and the
linear polarization $\perp$  to the symmetry directions
$\Sigma$ and $\Lambda$ were chosen in accordance with 
the experiment.\cite{tajima}

\begin{figure}
\vspace{-0.5cm}
\begin{center}
\begin{minipage}{8.0cm}
\begin{center}
\begin{minipage}{8.0cm}
\hspace{-0cm}
\psfig{file=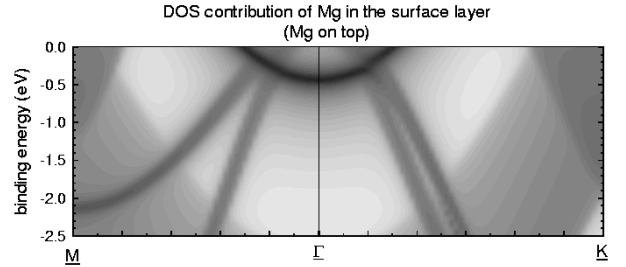,width=8.0cm,angle=-90}
\end{minipage}
\end{center}
\end{minipage}
\end{center}
\vspace{-0.0cm}
%\begin{figure}
\caption{% file: Mg_on_top-Mg-DOS-MGK-srf.ps
The DOS in the topmost layer 
(MgTS-case) {\it vs} binding energy
 and  $\mathbf{k}_{\parallel}$-vector along the 
 $\bar{\mathbf{\Gamma}} \bar{\mathbf{M}}$
($\Sigma$) and $\bar{\mathbf{\Gamma}} \bar{\mathbf{K}}$
 ($\Lambda$) symmetry directions. The DOS is measured by the darkness 
 on a logarithmic grey-scale.
The step-like surface potential barrier  is at a distance 
$d$=0.99~\AA\ from the
topmost layer.
Notice the Mg-derived SFS at $\approx$-0.5~eV at $\bar{\mathbf{\Gamma}}$. 
\label{MgB2DOSfig1}}
\end{figure}

In Fig.~\ref{MgB2DOSfig1} we show the DOS of the surface layer in
\vspace{0.0cm}
\noindent
 the case of a MgTS.
The quasi parabolic feature with a minimum around \mbox{-0.5~eV} on the
$\bar{\mathbf{\Gamma}}$ point is clearly a SFS, lying in a gap in the
$k_z$-projected bulk band structure \cite{silkin}.
It is $s$, $p_z$, $d_{z^2}$ in character, in a proportion roughly
3:2:1  at the $\bar{\mathbf{\Gamma}}$ point. Its characteristic decay
length is 2.6~\AA, whereas the distance between the Mg and B planes is
1.762~\AA. In Fig.~\ref{MgB2DOSfig1}, the dispersion of bulk bands is
also visible. The narrower stripes correspond to the quasi
2-dimensional $\sigma$ bonded p$_{x,y}$-bands, while the broad parabolic tape 
corresponds to the $\pi$ bonded p$_z$-band.

\begin{figure}
\vspace{-0.0cm}
\begin{center}
 \begin{minipage}{8.5cm}
\begin{center}
\begin{minipage}{8.5cm}
\hspace{-0cm}
\psfig{file=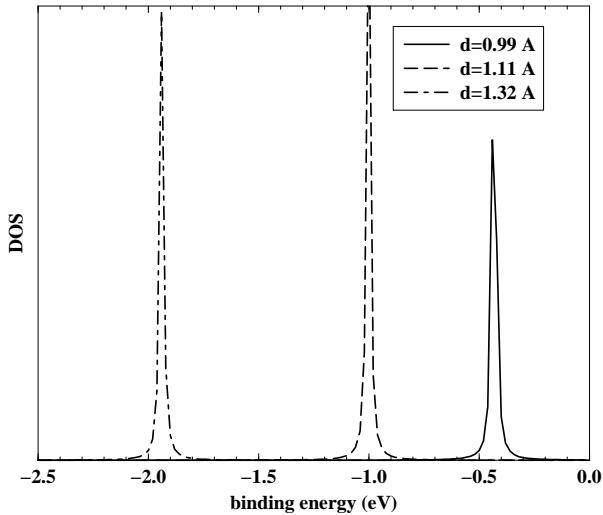,width=8.0cm}
\end{minipage}
\end{center}
\end{minipage}
\end{center}
\vspace{-0.0cm}
\caption{
% file: Mg_on_top-gamma.eps
The DOS of the
surface layer in the case of a MgTS as a function of the binding
energy, for $\mathbf{{k}}_{\parallel}=\bar{\mathbf{\Gamma}}$.
The different curves were calculated with 
step-like surface potential barrier placed at various 
distances $d$ (given in~\AA) from the surface. The imaginary part of the
self-energy was set to $\mathit{Im}\Sigma$=\mbox{-10~meV} in order to
sharpen the peaks. 
\label{fig:Mgpeaks}}
\end{figure}

\begin{figure}
\hspace{-0cm}
\begin{center}
\begin{minipage}{8.0cm}
\begin{center}
\begin{minipage}{8cm}
\hspace{-0cm}
\psfig{file=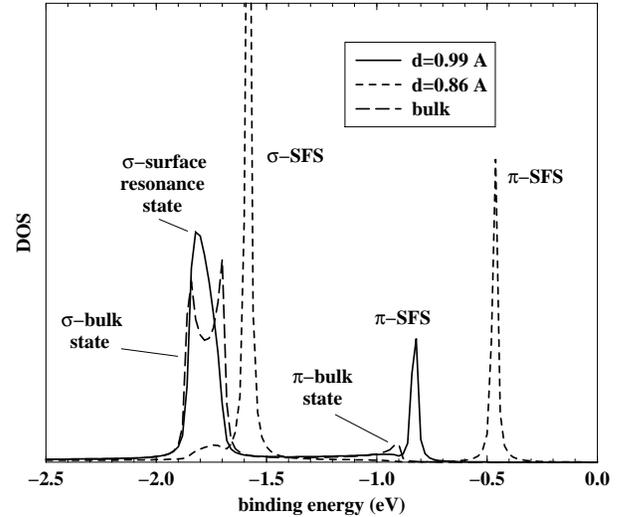,width=8.0cm}
\end{minipage}
\end{center}
\end{minipage}
\end{center}
\caption{
% file: B_on_top-k=436-bis.eps
The DOS of the surface layer in the BTS case 
as a function of the binding energy for
$\mathbf{k}_{\parallel}=\mathrm{0.7\cdot}\bar{\mathbf{M}}$.
The curves were calculated using a
step-like surface potential barrier placed at
different distances $d$ (given in~\AA) from the topmost layer.
$\mathit{Im}\Sigma$ as in Fig.~2.
\label{fig:Bpeaks}}
\end{figure}

SFS energies are much more sensitive to the values of the surface
potential barrier positions $d$ than to the work functions~$\phi$.
Placing the barrier at a larger $d$ value
results in a downward energy shift (cf.\ Figs.~\ref{fig:Mgpeaks} and \ref{fig:Bpeaks}). Thus, the
situation depicted 
in Ref.\ 2, where the
energy of the SFS of a MgTS is \mbox{$\approx$~-2~eV} at 
$\bar{\mathbf{\Gamma}}$, can be achieved with a surface potential
barrier placed at $d$=1.32~\AA\ above the topmost Mg layer.
A different situation occurs in the BTS-case (cf.\ 
Fig.~\ref{fig:Bpeaks}). Here the result of Ref.~2  
is reproduced by a smaller $d$-value than our empirical one. This 
is probably due to the in-plane B-B distance which is 57\%
smaller than the Mg-Mg one, so that in a BTS {\em slab} calculation, the
image plane position could have been chosen nearer to the surface than 
in the MgTS.
The SFS which we found for a BTS in the considered energy
range, runs parallel to the $p_z$ broad band  near its
upper edge. It is depicted by the solid line
peak at a binding energy $\approx$~-0.8~eV in Fig.~\ref{fig:Bpeaks} 
calculated at
$\mathbf{k_\parallel}=\mathrm{0.7\cdot}\bar{\mathbf{M}}$.
It decays towards 
the bulk 
with a characteristic length of 9.8~\AA \ and exhibits p$_z$ symmetry.
If steps are present on the real surface, 
yielding islands of Mg- and
BTS, a {\em single} effective 
\vspace{-1.3cm}
\vspace{-0.0cm}
\noindent
(averaged) $d$ value  could be justified
to analyze the ARPES data.
In view to have either a MgTS a BTS (or even both if we have a
step-like surface), we calculated the ARPES spectra shown in Figs.~4
and~6.

We would like to stress that
the crucial effective $d$ parameter must be fitted from experimental
data, and the only available data  \cite{tajima} is
shown in Fig.~\ref{fig:tajima}.
%In Fig.~\ref{fig:tajima}, the second derivative of the measured
%spectra is shown. 
Nice agreement with our calculations is
obtained, if one superposes Figs.~4 and~6, suggesting that the
measured sample presented both Mg- and BTS.
According to our calculations the feature centered at $\approx$0.4~eV at
$\bar{\mathbf{\Gamma}}$ is then
to be assigned to a $\pi$-bonded Mg derived SFS in the case of MgTS.
The parabolic feature centered at $\approx$2.7~eV is most likely, in our opinion,
to be assigned to a $\pi$-bonded B $2p_z$-derived SFS arising from a BTS, 
in accordance
with the calculations in Ref.~2.
However, we admit that from the presented experimental data, an
alternative assignment of this parabolic feature to the bulk B 2p$_z$
derived $\pi$-band edge cannot be completely excluded. 
An analogous consideration holds for the B $\sigma$ derived narrow
bulk stripes, where a surface resonance state is found with our 
$d$ parameter value (cf.\ Fig.~3).
Further ARPES experiments at variable photon energies should be
performed to solve the full assignment problem.
We emphasize that our interpretation mainly in terms of SFS does not
reduce the importance of Ref.\ 1 which is the first experimental
verification of the correct description of the electronic structure of
MgB$_2$  by the LDA.
This is very  important with respect to some times discussed strong 
many-body effects (correlation or  non-adiabaticity) which would cause 
\cite{alexandrov,pietronero}) 
 strong renormalizations of 
bulk and SFS as well. 
If strong correlation effects were present, they should
\noindent
\begin{figure}
\vspace{-0.0cm}
\begin{center}
\begin{minipage}{8.0cm}
\begin{center}
\begin{minipage}{8cm}
\hspace{-0cm}
\psfig{file=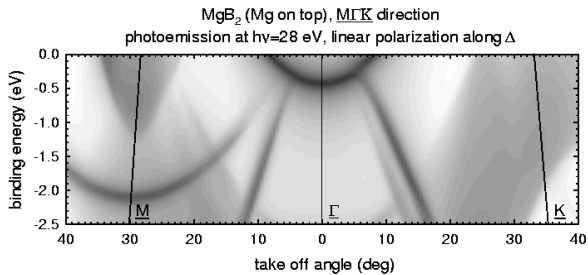,width=8cm}
\end{minipage}
\end{center}
\end{minipage}
\end{center}
\vspace{-0.0cm}
\caption{Theoretical ARPES energy distribution curves taken along $\Sigma$ 
(left panel) and $\Lambda$ (right panel) symmetry directions for a 
MgTS.  
The intensity is measured by the darkness.
\label{fig:Mgphoto}}
\end{figure}
\vspace{-1cm}
\begin{figure}
\hspace{-0cm}
\begin{center}
\begin{minipage}{8.0cm}
\begin{center}
\begin{minipage}{8cm}
\hspace{-0cm}
\psfig{file=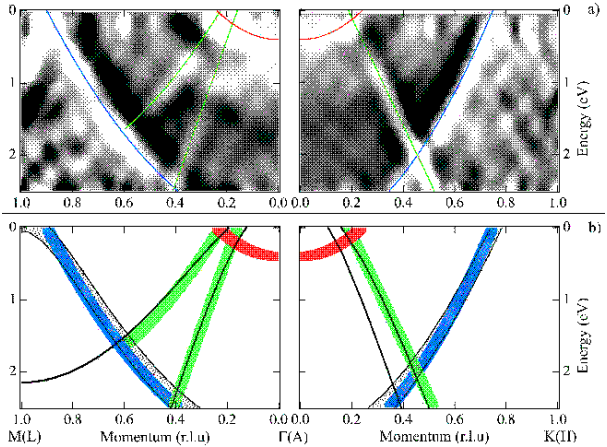,width=8.0cm}
\end{minipage}
\end{center}
\end{minipage}
\end{center}
\vspace{0.5cm}
\caption{% file: B_on_top-MGK.ps
((Color), after Uchiyama {\it et al.}[1])
Experimental ARPES data (upper panel). 
The intensity is measured by the brightness 
obtained from the second derivative of the ARPES spectra. 
The assignment of {\it one} SFS (red) and bulk states ($\pi$ (blue) and 
$\sigma$ (green)) proposed in Ref.\ 1 (lower panel).
\label{fig:tajima}}
\end{figure}
\vspace{-0.5cm}  
\begin{figure}
\hspace{-0cm}
\begin{center}
\begin{minipage}{8.0cm}
\begin{center}
\begin{minipage}{8cm}
\hspace{-0cm}
\psfig{file=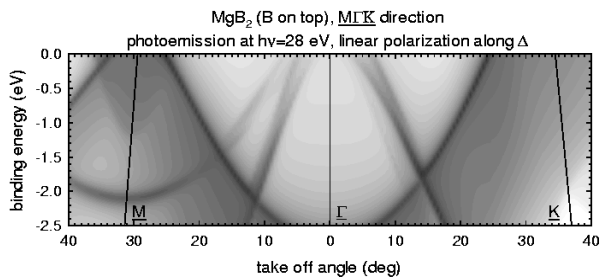,width=8.cm}
\end{minipage}
\end{center}
\end{minipage}
\end{center}
\vspace{-.0cm}
\caption{The same as in Fig.\ 4 for a BTS.
\label{fig:Bphoto}}
\end{figure}
 \noindent
  be observed
first in the SFS, because the screening of Coulomb interactions is
weaker near the surface rather than in the bulk.
Taking into account the perfect agreement of LDA with the ARPES data,
we conclude that such many-body effects should be of minor importance.
With respect to surface superconductivity to be discussed 
below, the experimental determination of the electron-phonon (el-ph) coupling 
constant $\lambda$ from the temperature dependence of the ARPES spectra 
like in Be would be of great interest \cite{bala,hengsberger}.

%\section{Discussion}
\par Now we address  possible consequences
of the SFS
with respect to physical properties of various samples with emphasis 
on superconductivity and micro-powder samples. Since for typical 
samples
the size in $c$-direction is about 10 nm only, about one quarter  
of the volume is affected by the presence of SFS.
Since for MgTS the SFS shown in Fig.\ 1 is derived from orbitals 
with cylindrical symmetry and there is no contribution from orbitals with 
 $p_{x,y}$ character,
a very 
weak local el-ph interaction is expected. Thus it acts as a bad or non- 
superconducting layer in the sense of the proximity effect. Due to the 
interaction with the bulk states the resulting 
superconductivity near  a MgTS will be weakened even for a clean 
MgTS, at variance with the view put forward in Ref.~3.

 Thus the 
presence of such SFS might explain the absence of 
a third upper critical field $H_{c3}$ \cite{lima} 
and the {\it local} suppression 
of superconductivity in relatively weak external magnetic field slightly 
above 2 Tesla  
\cite{budkoaniso,simon,papavassiliou}. This was 
regarded as evidence for a small value of $H^c_{c2}$, 
i.e.\ a big anisotropy $\gamma_H \sim $ 6 to 9. Anyhow, 
this would be in sharp contrast to the anisotropy caused by 
the Fermi velocities $\gamma^\sigma_v \sim$ 6 to 7 which yields 
$\gamma^\sigma_H\sim $4.2 to 5.5 as an upper limit \cite{langmann91}
derived for the strongly interacting subsystem of $\sigma$-holes. \cite{fuchs}
Due to the two-band character of MgB$_2$ \cite{fuchs,shulga,rosner} 
which involves  also 
the weakly interacting nearly isotropic 
$\pi$-electrons the total $\gamma_H$ is further reduced 
till below 4 to 5, 
provided any  order parameter anisotropy of the two different 
gaps can be neglected\cite{rosner,choi}. 
A moderate $\gamma_H\sim 3$ is in accord with available single 
crystal data.\cite{fuchs}

The interplay of superconducting  bulk and SFS subsystems
considered here 
contrasts the approach adopted in Ref.\ \onlinecite{bascones}
where due to symmetry breaking at the surface mixing and pair breaking 
of $\sigma$- 
and $\pi$- states in the sense of a two-band model occurs. 
Notably, the absence or presence of  superconducting SFS
may affect the surface pinning. Bulk
vortices will be attracted or repelled, respectively. 
This might  explain the strong variations of pinning properties
after applying pressure, which might affect the interfaces of the 
grains.
Furthermore the presence of surface superconductivity on SFS will 
affect the tunneling data.
In the context of frequently considered 
3D multi-band models for MgB$_2$ it is important 
to make a correct assignment of the observed superconducting gaps. 
In particular, in view of the ARPES interpretation 
we suggested above, 
the gap at about 5~meV observed by high-resolution (angle integrated) 
photoemission spectroscopy \cite{takahashi} might be ascribed to 
superconductivity on SFS.
  
Up to now two-dimensional (2D) electrons in metals were observed only in 
artificial layered structures\cite{Hsieh}.
Let us consider in short the new physics which can be gained by studying  
generic 2D electrons at the surface of a 3D superconductor. As most 
interesting we consider the possibility for enhancement of the
superconductivity of these 2D electrons due to increased  local DOS 
or softening of local phonon modes at the
surface. In this case in the framework of Ginzburg-Landau theory, for example, 
we have to introduce a 2D
superconducting order parameter $\Psi_{2D}$ in parallel to the bulk one 
$\Psi_{3D}$; additionally the
phenomenology could be complicated by necessity of taking into account pairing 
in  different and weakly
interacting bands. As a whole the surface enhancement of superconductivity 
could remind  the twinning plane
superconductivity observed in Sn and Nb.\cite{khlyustikov} In this case the 
dependence of 
magnetization as a function of temperature
and magnetic field can be similar to the fluctuation conductivity of the bulk. 
Roughly speaking, due to the
proximity effect 2D conductivity can create a ``frozen'' 3D fluctuation. A 
systematic study of fluctuation
magnetism in powdered samples with different grain sizes can give important 
hints in this direction. It will be
very informative to perform investigation of fluctuation 
magnetism\cite{Lascialfari} and to determine the mass
anisotropy of fluctuation Cooper pairs using MgB$_2$ microcrystals sintered 
under high pressure\cite{Filonenko}.

However it is more plausible that the influence of the surface is to 
reduce  the local superconductivity:
we have to take into account despite  the possible  
Mg or B terminations of the 
crystal, also the adsorption of various impurities {\it etc}\cite{kim}. 
In this case the influence of the surface can 
be traced out in the vortex pinning
in the superconducting state. Indeed, strong asymmetric hysteresis loops were 
observed after the grinding of a bulk
sample into fine powder\cite{Pissas}. 
Surface pinning was reported 
for a powder sample pressed into a pellet without sintering. 
\cite{Qin} Both 
possibilities, enhancement and suppression
of superconductivity yield additional surface pinning.  Finally, we 
argue that 2D electronic states
can give an important contribution to the current dependent correction of the work 
function of the superconducting metal, the so
called Bernoulli potential.

%\section{Conclusions}
\par We have shown that the available ARPES data can be well described 
by the LDA.  In particular, there is clear evidence
for the presence of at least two SFS. 
%Even more it seems that they dominated the measured ARPES intensities. 
Since the SFS depend very sensitively on the 
nature of the terminated layer, comparing our calculated intensities with 
the measured ones, we  suggest that the studied single crystal contained 
both types of terminated layers. Therefore the investigation of especially 
prepared surfaces with only one predominant surface layer is of interest.
In this context the theoretical study of steps between B and Mg
 terminated layers would be interesting.
Since the SFS are generated by the large gapped region in the BZ,
similar phenomena might be expected for other layered superconductors
such as NbSe$_2$ and even heavy fermion compounds URu$_2$Si$_2$ or 
borocarbides \cite{winzer}
where also no $H_{c3}$ has been observed. 
Taking into account all those possibilities mentioned above one can 
expect that  
comprehensive studies  of MgB$_2$ could
generate interesting physics of the surface/interface of superconductors.

\vspace{0.3cm}

%\section{Acknowledgments}
V.D.P.S.\ is grateful to A.N.~Yaresko and S.V.~Halilov who
respectively provided the LMTO and layer-KKR codes used in the
present study. We are  indepted to S.\ Tajima for 
providing us with 
the ARPES and other single crystal data 
prior to publication. We thank S.\ Tajima, H.\ Uchiyama, A.\ Kordyuk, 
G.\ Fuchs, S.\ Shulga, M.\ Richter, G.\ Paasch, J.\ Fink,  
H.\ Eschrig, C.\ Laubschat, and H.\ Rosner for  discussions. 
V.D.P.S.\ was supported  by the Deutsche Forschungsgemeinschaft 
(project La-655/7-1).

\end{document}